\documentclass[sigconf]{acmart}
\acmConference[Computation + Journalism Symposium (C+J '26)]{Computation + Journalism Symposium}{October 2--3, 2026}{Evanston, IL}
\setcopyright{none}
\acmISBN{}
\acmDOI{}
\makeatletter
\let\orig@footnotetextcopyrightpermission\footnotetextcopyrightpermission
\renewcommand\footnotetextcopyrightpermission[1]{%
  \orig@footnotetextcopyrightpermission{{\itshape Computation + Journalism Symposium (C+J '26), October 2--3, 2026, Evanston, IL.}}%
}
\makeatother

\usepackage{xurl}
\usepackage{listings}
\begin{document}

\title{Epstein Files Engine: Agentic Search for Investigative Journalism}

\author{Duy K. Nguyen}
\affiliation{
  \institution{The New York Times}
  \city{New York}
  \state{NY}
  \country{USA}}
\email{duy.nguyen@nytimes.com}

\author{Teresa Mondría Terol}
\authornote{Work done while at The New York Times.}
\affiliation{
  \institution{National Public Radio}
  \city{New York}
  \state{NY}
  \country{USA}}
\email{tmondriaterol@npr.org}

\author{Dylan Freedman}
\affiliation{
  \institution{The New York Times}
  \city{Washington}
  \state{DC}
  \country{USA}}
\email{dylan.freedman@nytimes.com}

\author{Zach Seward}
\affiliation{
  \institution{The New York Times}
  \city{New York}
  \state{NY}
  \country{USA}}
\email{zach.seward@nytimes.com}

\renewcommand{\shortauthors}{Duy K. Nguyen, Teresa Mondría Terol, Dylan Freedman and Zach Seward}
\begin{abstract}
On Jan.\ 30, 2026, the U.S. Department of Justice released a mixed-media collection concerning Jeffrey Epstein, including about three million pages of PDFs. We describe the Epstein Files Engine, an A.I.\ agent The New York Times deployed to investigate the files. The Engine translated reporter questions into Google BigQuery SQL queries across three corpora: Epstein-related releases, the Times's archive and external, Epstein-related news headlines. It used an LLM to plan queries and returned citation-rich answers a reporter could verify and trust. More than 100 journalists used the Engine, and it contributed to at least 20 published stories. We report how reporters queried it and describe Diff, our text-and-visual duplicate matching method that amplified novelty signals and allowed the Engine to surface genuinely new information. We argue that newsroom agents serve newsrooms best not as autonomous writers, but as interfaces to source material and institutional knowledge.
\end{abstract}

\keywords{computational journalism, document search, large language models, text-to-SQL, deduplication, newsroom tools, investigative reporting}

\maketitle

\section{Introduction}

On Jan.\ 30, 2026, the Department of Justice released a collection of records concerning Jeffrey Epstein, the former financier who died in prison six years earlier on federal sex-trafficking charges~\cite{doj2026epstein}. An internal Times inventory put the release at roughly three million pages of PDFs, 1.1 million emails, 200,000 images and more than 1,000 audio and video clips. With a printed stack approaching the Empire State Building in height~\cite{nytEpstein}, it was both too large to read exhaustively and too varied for keyword search to organize.

Retrieval, though, was only part of the problem. At what we coin the \emph{first mile}, a reporter onboarded to the Epstein beat begins with a hunch, and keyword search rewards the prior knowledge and experience they do not yet possess to turn that hunch into effective queries. At the \emph{last mile}, a relevant page or transcript snippet becomes a lead only once journalists have compared it against existing reports and reviewed the document where it originated.

The Times addressed these challenges with the Epstein Files Engine. Unlike traditional retrieval-augmented generation, the Engine dynamically translated a reporter's question in plain English into BigQuery SQL queries over three institutional corpora and answered with records and citations it retrieved, expediting both open-ended exploration and verification in a high-pressure news moment. More than 100 reporters used it for at least 20 published stories across politics, finance, sports, education and entertainment.

This paper makes two contributions. First, it reports a live newsroom deployment of such a grounded document-search agent during an active investigation and how reporters used it to produce leads. Second, it describes Diff, a multimodal duplicate matching method we used to amplify novelty signals in the release. Diff combines a semantic text embedding and a perceptual image fingerprint into a single vector representation optimized for BigQuery retrieval. Evaluated over its deployed decision threshold, not only does it reach high recall in exchange for low precision, many of its false positives reveal structural similarities meaningful to reporting, such as different redactions on the same document page, repeated forms and record genres.

\section{Related work}

The Epstein Files Engine belongs to a long line of document retrieval systems for reporting. DocumentCloud standardizes how newsrooms organize, annotate, search and publish primary-source documents~\cite{documentcloud}. Datashare and Google Pinpoint extract named entities from, index and empower newsrooms to collaboratively analyze huge collections such as the Panama, Paradise and Pandora Papers~\cite{icijDatashare, googlePinpoint}. Semantra enables local semantic search and query refinement over text corpora without generating answers~\cite{semantra}. Hagar, Diakopoulos and Gilbert show how underresourced newsrooms can use small language models to host their own retrieval-augmented generation pipelines for investigative document search~\cite{hagar2025rag}.

To the extent that these systems play a role in the making of a news story, they share a sibling relationship with ideation and discovery tools for journalism. AngleKindling uses LLMs to frame a press release in various ways and help reporters avoid pushing the narratives and agendas it espouses~\cite{petridis2023anglekindling}. Veerbeek and Diakopoulos confirm viability of a multi-agent system to generate investigative tip sheets over structured data~\cite{veerbeek2024}. Broussard, a decade ago, developed an expert system that generated interactive visualizations from public school data, surfacing opportunities for reporting and open-ended exploration~\cite{broussard2015artificial}. The Engine takes inspiration from this body of work, employing an LLM amplified with Diff-based novelty signals to catalyze lead discovery.

On the audience side of computational journalism~\cite{cohen2011computational}, applications that use RAG over an institutional corpus have risen in popularity. The Financial Times' ``Ask FT'' and The Washington Post's ``Ask The Post AI'' generate answers to reader questions from their respective reporting archives~\cite{askFT, askThePost}. The Times's Wirecutter search likewise uses A.I.\ to surface product recommendations from its service journalism ~\cite{wirecutterSearch}. These applications exemplify Fang's theory of the ``intimacy dividend,'' in which people see chatbots as a socially forgiving interface to ask questions they might hesitate to ask others~\cite{fang2026intimacy}. If previous work on lead discovery empirically confirms newsroom-side existence of this dividend, as we will show, The Times's deployment of the Engine definitively grounds that existence in the journalism it enabled.

Lastly, text-to-SQL generation over large schemas is a known research problem. Benchmarks such as Spider, BEAVER and EntSQL track SQL generation performance over enterprise workflows~\cite{lei2024spider2, chen2024beaver, liao2026entsql}, but part of the Engine's innovation is its application of SQL generation to newsroom workflows at the scale of the biggest newsroom in the U.S.~\cite{hunter2026hasnyt}. Diff itself sits within a lineage of multimodal representations. If trained, multimodal embeddings such as CLIP, SIGLIP and ImageBind show the power of universal representations~\cite{radford2021clip, zhai2023sigmoid, girdhar2023imagebind}, Diff simply concatenates two modality signals whose strengths and weaknesses journalists can inspect. What journalists can inspect, they can learn to trust.

\section{System design}

The Epstein Files Engine answered reporter questions by translating them into BigQuery SQL queries over three corpora. Its interface ran as an agent on the Times's self-hosted version of LibreChat, an open-source chat platform with agent and tool support via the Model Context Protocol~\cite{librechat, mcpIntro2026}.

\subsection{The Engine's underlying data}

The first corpus was the Jan.\ 30 release, plus earlier Epstein-related releases that The Times acquired, preprocessed and persisted in a set of BigQuery tables. To address photographs and screenshots embedded within document pages, we first extracted them with a custom image detection model, then used a vision-language model to write tags and captions of each detected image into a companion table. Doing so allowed visual assets ``trapped'' within PDFs to be queryable by the Engine. The second corpus was the Times's own reporting archive, including our prior coverage on Jeffrey Epstein. The third was an internal feed of headlines mentioning Epstein from other news organizations, which the Engine used to check whether outside coverage was already pursuing the same thread. (The Times does not use these headlines for any other purpose, including training LLMs.)

What made the Engine useful was its ability to both weave insights across these three different corpora and suggest, in mere minutes, leads that would otherwise take days, weeks or even months to surface. Its programming accommodated varying levels of verbosity in the user query. In the simplest scenario, a reporter assigned to cover an individual appearing in the Epstein Files only needed to enter their name. The Engine understood that they were not asking only for what the documents said about that name, but \textit{what was actually new about it}, proxied by what The Times itself and external headlines on Epstein had already reported on.

\subsection{The model as query planner}

Our design emphasizes SQL query planning over answer generation. At the Engine's heart, an LLM translates a reporter question into queries executed over the corpora and synthesizes a report whose source of truth remains the records it cites. This separates it from a generic retrieval-augmented chatbot, which fixes database queries and aims to generate an authoritative and final answer.

Figure~\ref{fig:query} shows an anonymized version of the queries the Engine generated for one reporter's question, one per corpus. The Engine ran them separately and assembled the returned records into a single citation-rich report.

\begin{figure}[t]
\begin{lstlisting}[language=SQL]
-- Reporter: "What do the files say about <subject>,
--   and what has not already been reported on them?"

-- (1) New release pages, dropping re-releases Diff flagged
SELECT p.doc_id, p.document, p.page, p.text, p.source_url
FROM   release.pages p
LEFT JOIN diff.duplicates d ON d.page_id = p.doc_id
WHERE  p.tranche IN ( /* the DOJ release batches */ )
  AND (LOWER(p.text) LIKE '<subject>'
       OR  LOWER(p.text) LIKE '<name variant>')
  AND  d.is_duplicate IS NOT TRUE;

-- (2) What the newsroom has already published
SELECT headline, url, published_at
FROM   archive.published_assets
WHERE  LOWER(article_body) LIKE '<subject>'
ORDER  BY published_at DESC;

-- (3) Whether other news organizations are chasing the same thread
SELECT title, url
FROM   external_epstein_headlines
WHERE  LOWER(title) LIKE '<subject>';
\end{lstlisting}
\caption{A reporter question and anonymized versions of the BigQuery queries the Engine generated for it. The queries search the Epstein document corpus, the Times archive and external, Epstein-related headlines. The Engine then assembles returned records into one cited report.}
\Description{An anonymized SQL listing showing three generated queries: one searches Epstein document pages while excluding Diff-flagged duplicates, one searches prior Times coverage and one searches external, Epstein-related headlines.}
\label{fig:query}
\end{figure}

\subsection{Diff's deduplication mechanism}

A document corpus is not simply a set of distinct pages. Tranches re-release the same material with new Bates numbers. Scanned duplicates may differ from their originals in image noise and tilt angle. The Times's Epstein corpus also included earlier Epstein disclosures, such as congressional releases and court files. Surfacing what is genuinely new, relative to prior releases and other pages in the same release, was a condition for journalistic impact.

We thus devised a page-level duplicate matching method called Diff. It represents each page as a 449-dimension composite vector: 384 dimensions of embedding generated from Google's \texttt{gemini-embedding-001} model on the semantic-similarity task against text extracted (with optical character recognition, if needed) from that page; 64 dimensions obtained from vectorizing an $8\times8$ difference hash of the page, rendered as an image; and a single bias term. A difference hash of an image is calculated by comparing brightness values between adjacent pixels~\cite{hoytDhash}. It allows two scans of the same page to match even when they differ in resolution or brightness.

Formally, we encode each page $p$ as a text embedding $\mathbf{t}(p)\in\mathbb{R}^{384}$ and a binary perceptual hash $\mathbf{h}(p)\in\{0,1\}^{64}$. Each nonzero modality is independently $\ell_2$-normalized to give equal weights to semantic and visual signals. Where a modality is absent (for example, when the ``page'' is in fact a snippet of video or audio transcript, or when a PDF page is fully redacted and thus contains no extractable text), it is represented as $\mathbf{0}$:
\[
\hat{\mathbf{t}}(p)=\mathbf{t}(p)/\lVert\mathbf{t}(p)\rVert_2,\qquad
\hat{\mathbf{h}}(p)=\mathbf{h}(p)/\lVert\mathbf{h}(p)\rVert_2.
\]
To prevent all-zero pages (for example, an all-white or blank PDF page) from producing undefined similarity values, we append a constant bias term to the composite representation,
\[
\mathbf{v}(p)=\big[\,\hat{\mathbf{t}}(p)\ \Vert\ \hat{\mathbf{h}}(p)\ \Vert\ 1\,\big]\in\mathbb{R}^{449}.
\]
Duplicate detection is performed over nearest-neighbor search using cosine similarity over $\mathbf{v}$,
\[
s(p,q)=
\frac{\hat{\mathbf{t}}(p)\cdot\hat{\mathbf{t}}(q)+\hat{\mathbf{h}}(p)\cdot\hat{\mathbf{h}}(q)+1}
{\lVert\mathbf{v}(p)\rVert_2\,\lVert\mathbf{v}(q)\rVert_2}.
\]
A page that carries both signals has $\lVert\mathbf{v}\rVert_2=\sqrt{3}$. Letting $s_{\text{text}}$ and $s_{\text{vis}}$ denote the text and visual dot products, the score reduces to
\[
s(p,q)=\frac{s_{\text{text}}(p,q)+s_{\text{vis}}(p,q)+1}{3}.
\]

Note our choice of cosine instead of the usual Hamming distance to compare visual hashes. If two binary hashes have both $k$ set bits and Hamming distance $H$, their cosine similarity (also the dot product of their normalized versions) is $1-H/(2k)$. When the number of set bits differs, cosine is no longer determined by Hamming distance alone, though it still rewards shared adjacent-brightness comparisons while normalizing for bit count.\footnote{For hashes $x$ and $y$, let $\hat{x}$ and $\hat{y}$ be their normalized versions, $k_x$ and $k_y$ be their set-bit counts and $c=x\cdot y$ the number of shared set bits. Their Hamming distance is $H=(k_x-c)+(k_y-c)=k_x+k_y-2c$, so $\hat{x}\cdot\hat{y}=\cos(x,y)=c/\sqrt{k_xk_y}=(k_x+k_y-H)/(2\sqrt{k_xk_y})$. When $k_x=k_y=k$, this reduces to $1-H/(2k)$.} Folding text and visual signals into a single vector enables the Engine to run a single BigQuery \texttt{VECTOR\_SEARCH} over the page corpus, simplifying deployment, maintenance and agentic use in such a high-pressure investigation~\cite{googleBigQueryVectorSearch}. Text embeddings catch semantic re-releases where layout changed, and the visual hash catches scanned duplicates where optical character recognition garbled the text or where there was little text to begin with. Both signals are needed for a release that mixes born-digital filings with photocopied exhibits.

Finally, let $\tau$ be the duplicate-detection threshold. We flag pages $p$ and $q$ as duplicates when $s(p,q) \ge \tau$. Concatenating text and visual signals, as opposed to a trained fused representation such as CLIP, enabled the Engine to isolate the contribution of each signal to the duplicate decision in query and allowed its developers to manually inspect Engine results and tune $\tau$ in real time. We set $\tau = 0.92$ as the deployed threshold.

\section{Deployment and use}

A local prototype took its first query on Feb.\ 1, 2026, two days after the release. Over the next week, the newsroom's A.I.\ Initiatives team ran reporter questions through it and pasted cited reports into a shared Google Doc (which has since hit its 1 million-character limit). The Engine itself went live to reporters on Feb.\ 9, turning that manual relay into self-serve access. We presented the Engine to Times reporters with three explicit cautions:

\begin{enumerate}
  \item statistical claims were not to be trusted;
  \item Engine reports were to be treated as fallible; and
  \item findings were to be verified against the cited documents and always accompanied with original reporting.
\end{enumerate}

Between launch and mid-June, its server logged more than 100 journalists posing 4,500 questions to the agent. A subset of those questions developed into  written reports that carried document citations and a comparison to prior coverage, and at least 20 of those reports contributed to published articles. Qualitatively, we characterize the kinds of questions reporters asked from these reports and from the requests they posted to a dedicated Slack channel. The categories below are not a classification of all messages with the Engine, which are confidential.

\subsection{What reporters asked}

One common request was a profile of an individual: given a name, the Engine returned what the documents contained about them, who else appeared alongside them and what was new relative to the Times's and external Epstein headline coverage. Between the local prototype and the deployed service, more than 60 such named-entity reports were produced in the first weeks of the investigation.

After being trained to use the Engine, reporters frequently turned to it precisely for its ability to reason across corpora --- to contextualize what the files said about a person within what The Times and external headlines had already published to determine newness (and sometimes newsworthiness) of their findings. To accommodate this line of inquiry, we instructed the Engine to include a dedicated section for existing coverage in its reports. This single use case united the first- and last-mile problems, where the chat interface allowed reporters to come with the most nebulous ideas and, in minutes, walk away with a solid lead that launched their reporting.

In other requests, financial-trail queries looked for transactions not yet reported, relationship-mapping queries surfaced the network around a name, queries constructed timelines of contact between two people and queries surfaced euphemisms and coded language in Epstein's communications. In one forensic use, the Engine attempted to identify the author of an anonymous document by extrapolating their writing style and surfaced similar writing where their name was not redacted. Reporters also probed its reliability, asking whether an irrelevant result meant an attempt at fabrication or reading beneath a redaction. It was neither, since the Engine was not instructed to recover redacted text and said so.

\subsection{From query to published article}

There was a consistent path from hunch to story. After a reporter posed a question, the Engine returned a report grounded in document citations and a comparison to coverage from existing Times stories and external headlines on Epstein. The reporter then vetted the cited material against the original documents and built on it with original reporting into a story. Every user was trained to treat Engine output as fallible and verify it against original documents.

Among the roughly 60 targeted reports, at least 20 contributed to published articles spanning politics, finance, sports, education and entertainment. They included Ghislaine Maxwell's role in the Clintons' circle, the departure of Goldman Sachs's top lawyer after her ties to Epstein surfaced, Leon Black's reliance on Epstein, the Harvard figures eager to align themselves with him, his childhood and untold details surrounding his death~\cite{nytMaxwellCGI, nytRuemmler, nytBlack, nytHarvard, nytChildhood, nytDeath}. In one example, after a reporter supplied questions about a fixture of British politics, the Engine answered each against cited documents, and the story was published days later~\cite{nytMandelson}. Many queries that did not become stories concerned living people and unproven allegations, and The Times's verification standard worked precisely because most leads did not survive it.

\subsection{Validating Diff}
\label{sec:validation}

We retrospectively validate Diff against human judgment on The Times's inventory of the Jan.\ 30 release, which alone includes 2,683,926 pages of PDF documents, along with video and audio transcripts. We map each page to its nearest composite neighbor and partition the population into four bands on the basis of composite similarity. We draw a stratified sample of 400 pairs across the bands, oversampling the bands around the deployed $\tau=0.92$ cutoff (Table~\ref{tab:strata}). This sample size keeps expert annotation feasible. We draw 95\% confidence intervals from a stratified bootstrap.

\begin{table}
  \caption{Stratified sample over Diff's per-page nearest-neighbor population, partitioned using four bands on the basis of composite similarity. We intentionally oversample the bands around the deployed threshold $\tau=0.92$. Pairs in high and medium bands are flagged as duplicates; pairs in low and lowest bands are not.}
  \label{tab:strata}
  \small
  \begin{tabular}{@{}llrr@{}}
    \toprule
    Band & Similarity & Population & Sampled \\
    \midrule
    High    & $\geq 0.97$    & 82,417    & 60  \\
    Medium  & $0.92$--$0.97$ & 547,899   & 150 \\
    \midrule
    Low     & $0.80$--$0.92$ & 2,052,114 & 150 \\
    Lowest  & $0.74$--$0.80$ & 1,496     & 40  \\
    \midrule
    Total   &                & 2,683,926 & 400 \\
    \bottomrule
  \end{tabular}
\end{table}

We label each pair \textit{exact duplicates}, \textit{near-duplicates} or \textit{distinct} under a rubric that reflects how Times journalists identify duplicates. Two pages are distinct if they clearly come from different underlying documents. A re-release under a new Bates number is exact, but if the pair is the same page, just redacted differently, it is a near-duplicate. Two records on a shared form template and two unrelated documents carrying the same boilerplate footer are both distinct. A blank or fully-redacted page whose content cannot be observed is marked unsure and excluded. This rubric is explicit enough to reproduce the labels exactly. We thus report one annotator's labels.

At the deployed threshold $\tau=0.92$, Diff reaches a population-weighted precision of $0.28$ and a recall of $0.86$ (Table~\ref{tab:sweep}). Among pairs labeled distinct, the true-duplicate (false negative) rate is only 1.4\%, meaning a small percentage of genuine duplicates was not caught. Precision is low for a reason. In the medium ($0.92$--$0.97$) band, we estimate that 22.5\% (95\% CI $[0.16, 0.31]$) of the 547,899 predicted exact duplicates are true duplicates. (As later shown, many false positives are not random error but contain structural similarities journalistically useful in their own right.) Pairs marked unsure make up 18\% of the sampled population. Resolving them adversarially as exact duplicates or distinct moves precision within $[0.21, 0.43]$ and recall within $[0.59, 0.83]$.

As Table~\ref{tab:sweep} shows, no operating point is both precise and complete. A higher $\tau$ threshold trades recall for precision. Our deployed $0.92$ threshold, manually tuned in real time and in response to Times reporters' qualitative feedback on the Engine's overall answer quality, favors recall and tries to surface (and ignore) as many genuine duplicates as possible. It reflects an institutional need to separate signal from noise and identify genuinely new information in a high-stakes, breaking news environment.

\begin{table}
  \caption{Diff's duplicate decision at three decision thresholds. A pair is flagged as a duplicate when its composite cosine similarity is at least $\tau$, so a lower $\tau$ trades precision for recall. Values are population-weighted. At the deployed $\tau=0.92$, the 95\% bootstrap confidence intervals for precision, recall and $F_1$ are $[0.21, 0.35]$, $[0.69, 1.00]$ and $[0.34, 0.50]$.}
  \label{tab:sweep}
  \small
  \begin{tabular}{@{}lrrr@{}}
    \toprule
    Threshold $\tau$ & Prec. & Rec. & $F_1$ \\
    \midrule
    $0.97$ (strict)   & 0.64 & 0.26 & 0.37 \\
    $0.92$ (deployed) & 0.28 & 0.86 & 0.42 \\
    $0.80$ (loose)    & 0.08 & 1.00 & 0.14 \\
    \bottomrule
  \end{tabular}
\end{table}

\section{Findings and lessons}

Our clearest finding is that reporters used the Engine in ways its design anticipated but its creators did not script. That alone speaks to its versatility. In addition to archive comparisons, they also asked the Engine to dig for unreported financial transactions, reconstruct relationship networks and timelines of contacts, surface use of coded language between Epstein and his associates and analyze writing styles.

Times reporters, through our training and their previous experience with chatbots, naturally treated Engine's chat interface and its answers with skepticism. But grounding Engine reports in verifiable citations gave them something concrete to inspect and trust. In Fang's theory of the intimacy dividend, a chat interface is forgiving, drawing out questions audiences might hesitate to ask one another~\cite{fang2026intimacy}. The Engine shows the same dividend newsroom-side, in the exploratory and skeptical questions reporters asked it that they might not pose to a colleague or their editor, or know how to write as a SQL query, or possess prior knowledge about Epstein as a beat to draw connections across results.

\subsection{Limitations}
\label{sec:limitations}

Our contribution is primarily a record of newsroom deployment and the theoretical work behind it. We reconstruct and quantify use from journalist questions and written reports, and thus describe a floor on newsroom activity rather than a distribution. We did not perform a controlled study that withheld The Times's archive or external, Epstein-related headlines. That these corpora helped turn results into leads rests on qualitative descriptions of how reporters queried the Engine, how we instructed the Engine to shape its reports and the stories these reports enabled. We invite future research to build on our work and empirically test their efficacy.

Likewise, our validation of Diff happened after the news moment had passed. It is not an ablation study. We deployed the Engine with its composite representation and duplicate-decision threshold first to meet that news moment, tuned the latter in real time and did so manually. We did not obtain the $\tau=0.92$ value as a result of training a model against pre-registered ground truth.

We estimate precision and recall over a stratified sample of the nearest-neighbor population, but we did not compare the composite representation with its text-only or visual-only variant. In our validation, the ground truth asks whether the nearest composite neighbor of a page is a true duplicate, not whether every true duplicate in the corpus was found. We only seek to justify our decision to include Diff in the Engine's decision, and we invite future research to comprehensively benchmark it against alternatives.

To our knowledge, the Engine is the first documented effort at combining SQL generation and multimodal duplicate matching to a journalism project of this scale, but we are one newsroom, and this is one investigation. We report a system that worked in a specific deployment instead of a robust method across all journalistic uses.

\subsection{What near-duplicates reveal}
Deduplication treats near-duplicates as noise to suppress. Read another way, though, structural sameness has multiple uses depending on how coarsely it is measured — and Diff becomes a way to find things.

\begin{figure}
  \centering
  \small
  \setlength{\tabcolsep}{1pt}
  \begin{tabular}{@{}cc@{}}
    \multicolumn{2}{@{}l@{}}{\textbf{Same page, different redactions} ($s=0.9205$)} \\
    \includegraphics[width=.49\columnwidth]{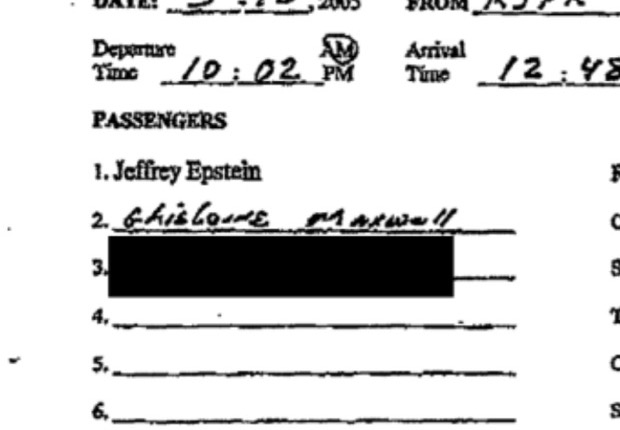} &
    \includegraphics[width=.49\columnwidth]{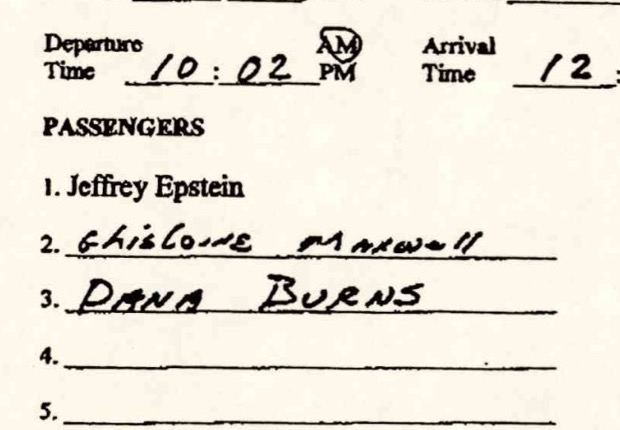} \\
    \multicolumn{2}{@{}l@{}}{\textbf{Same form template across dates} ($s=0.9705$)} \\
    \includegraphics[width=.49\columnwidth]{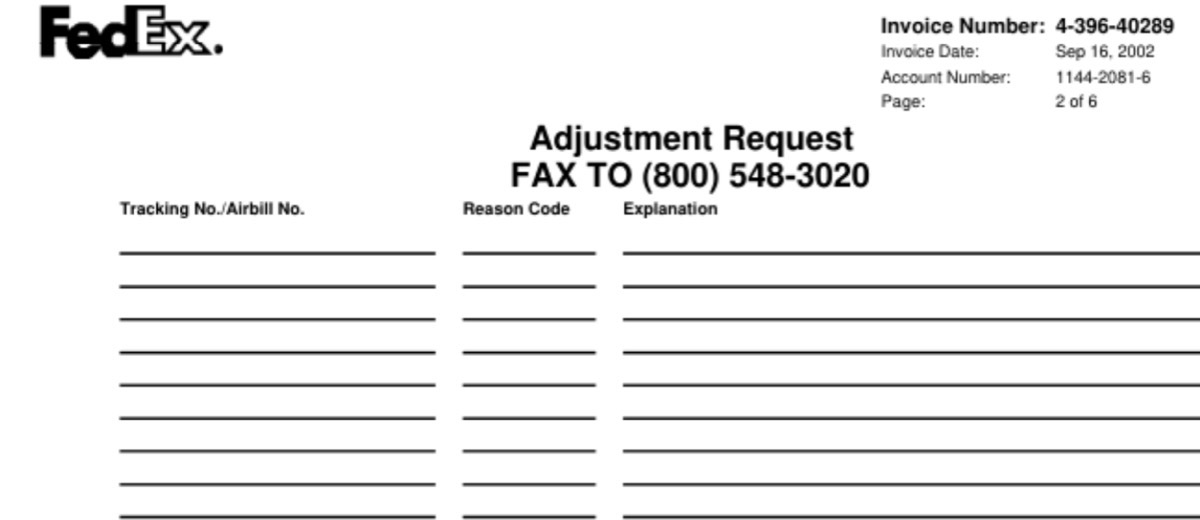} &
    \includegraphics[width=.49\columnwidth]{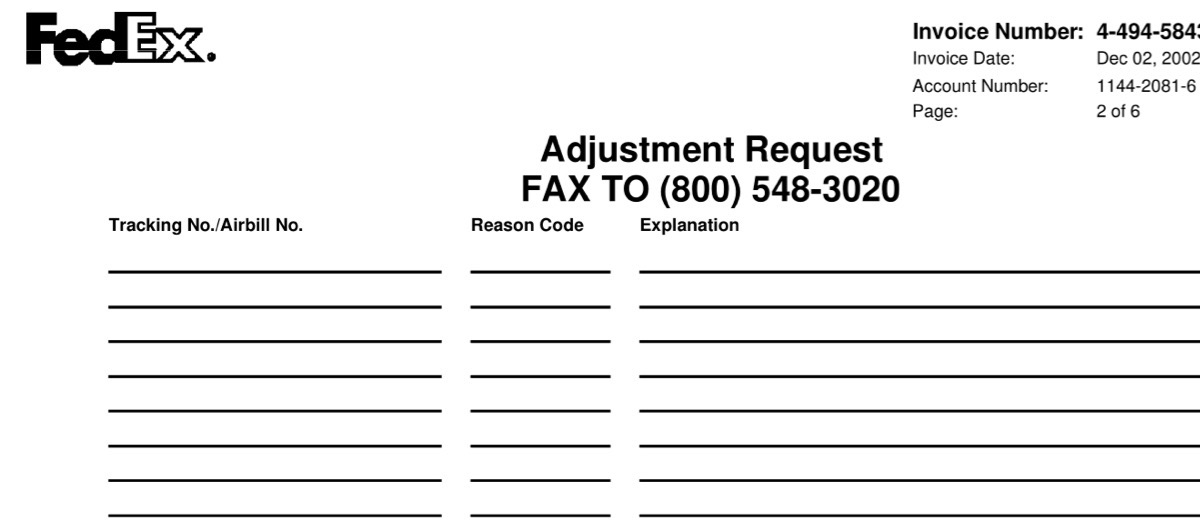} \\
    \multicolumn{2}{@{}l@{}}{\textbf{Different templates in one record genre} ($s=0.9134$)} \\
    \includegraphics[width=.49\columnwidth]{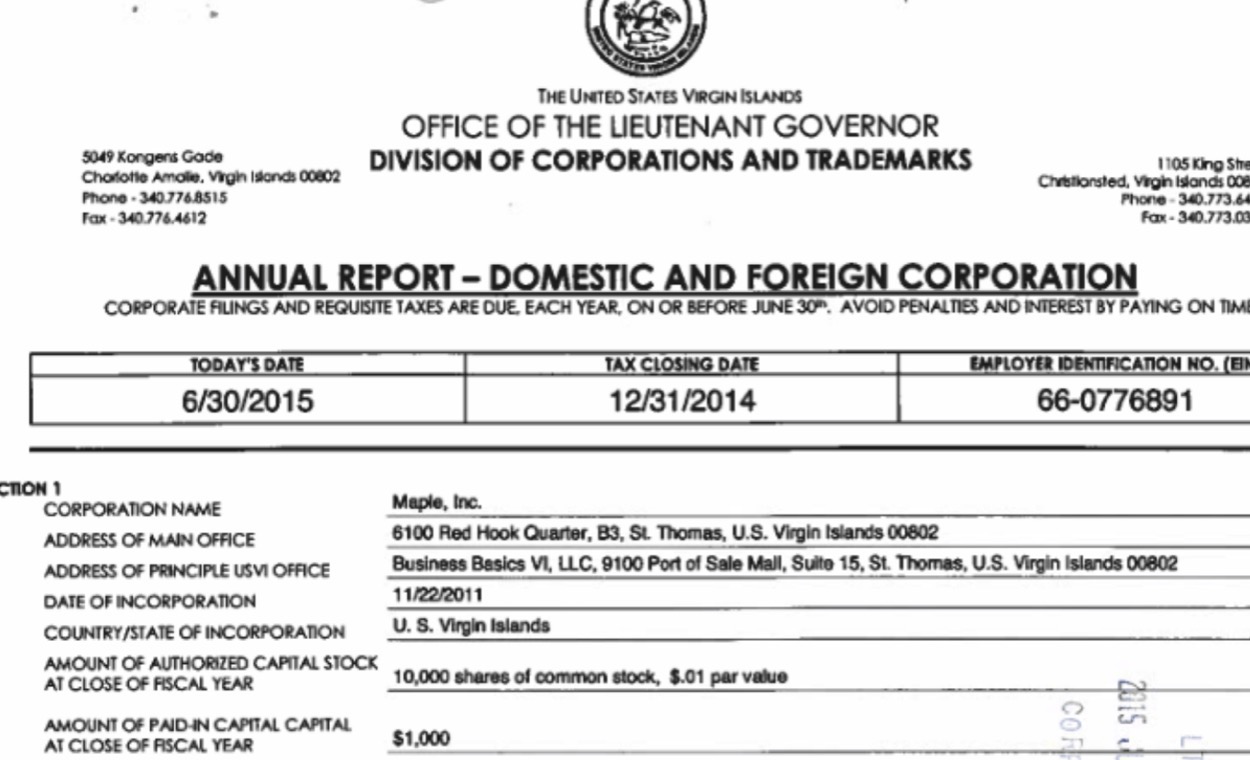} &
    \includegraphics[width=.49\columnwidth]{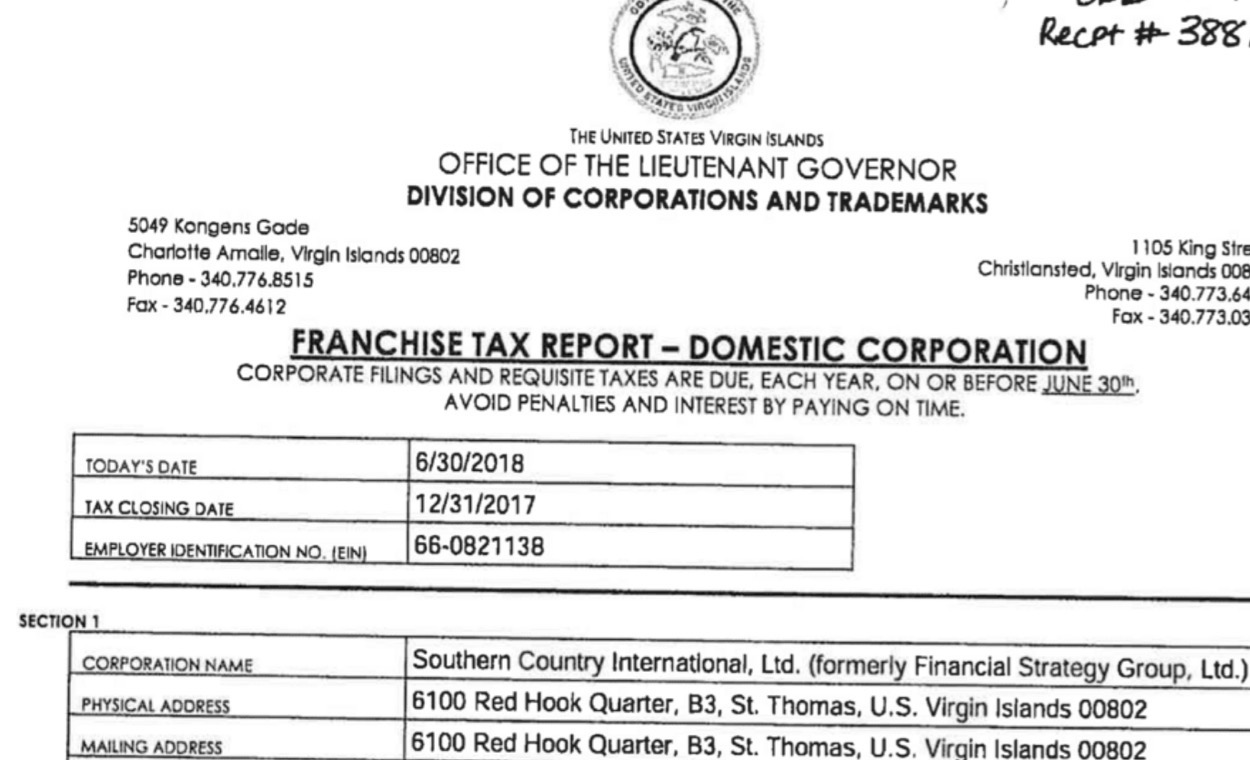} \\
  \end{tabular}
  \caption{Three resolutions of Diff's near-duplicate matches in the validation sample, showing why the same matches can also suggest reconstruction and genre-mapping uses.}
  \Description{Three pairs of document crops: a passenger manifest with and without a redaction over one passenger line; two FedEx adjustment-request forms with different invoice dates; and two U.S. Virgin Islands corporate records with related but different report templates.}
  \label{fig:resolution}
\end{figure}

Figure~\ref{fig:resolution} shows three resolutions of Diff's near-duplicate matches in our labeled validation sample. At the finest resolution, the same page is released twice but \emph{under different redaction masks}, uncovering names and details that could otherwise get lost in the haystack. These pairs are both most useful and easiest to miss, since both their semantic and visual differences are subtle enough to push their composite similarities close to the decision threshold. One step coarser, a single form filled out across dates signifies a document series. Lining up filings tracks what changes between them and facilitates timeline reconstructions. Coarser still, different templates of the same document ``genre,'' such as tax and financial reports from the U.S. Virgin Islands, allow the Engine to cluster and surface similar records in query, gesturing reporters toward parts of the release they might not have ample prior knowledge to look for. This is solving the first mile in action.

\subsection{What generalizes}

We designed both the infrastructure and playbook behind the Epstein Files Engine with modularity and reusability in mind. Replace the Epstein files with any other primary-source document corpus, and the Engine can help reporters reason through them with Diff-based duplicate detection built in. We can also improve the self-serve relay with documentation that allows journalists to build their own Engines, effectively unlocking ``chat with your documents'' as an institutional capability. Again, unlike RAG, this way of chat allows dynamic contextualization across corpora and enables reporter verification and trust through query planning and citation.

Treating documents as structured data is by no means new; it was a principal vision behind DocumentCloud when it was founded in 2009~\cite{seward2009knight}. That the Engine operates under the same vision allows a newsroom to move beyond documents and create Engines that reason across \textit{datasets}, from census results and weather events to audience analytics and internal tooling metrics.

Thus, we claim a broader design stance: A newsroom agent is most useful as an interface to institutional memory and source material, not an autonomous writer. Its value lies not in knowing an authoritative, final answer, but in helping journalists pose better questions and weaving evidence across data silos into answers they can independently verify. It grants them analytical capabilities previously unimagined and then leaves judgment in their hands.

\begin{acks}
The authors thank Kirsten Danis, David Enrich and Matea Gold, whose editorial leadership turned the Jan.\ 30 release into a coordinated newsroom investigation, and the Times journalists who tested the system with their reporting. We thank Benjamin Koski, Andrew Chavez and the Interactive News Technology team for the acquisition, ingestion, auditing and infrastructure that made the release a searchable newsroom corpus, without which the Engine would not exist. We also thank our A.I.\ Initiatives colleagues James O'Toole, Juliana Castro Varón and Rubina Fillion, who extracted and enriched the release's non-text media, trained reporters to use the Engine and edited this paper line by line.
\end{acks}

\bibliographystyle{ACM-Reference-Format}
\bibliography{refs}

\end{document}